
\input harvmac
\noblackbox
%
\font\fourteenrm=cmr12 scaled \magstep1

\def\NPB#1#2#3{Nucl. Phys. {\bf B #1} (#2) #3}
\def\PLB#1#2#3{Phys. Lett.  {\bf B #1} (#2) #3}
\def\PR#1#2#3{Phys. Rev. {\bf  #1} (#2) #3}
\def\PRD#1#2#3{Phys. Rev. {\bf D #1} (#2) #3}
\def\PRL#1#2#3{Phys. Rev. Lett. {\bf  #1} (#2) #3}

\def\PRL#1#2#3{Phys. Rev. Lett. {\bf  #1} (#2) #3}

\def\RMP#1#2#3{Rev. Mod. Phys.  {\bf #1}  (#2)  #3}
\def\NC#1#2#3{Nouvo Cim.  {\bf #1}  (#2)  #3}
\def\fs #1{
{#1 \hskip -5pt/}
}
\def\fsm #1{
{#1 \hskip -7pt/}
}


\def\Lag{{\cal L}}

\def\Tr{{\rm Tr}}
\def\tr{{\rm tr}}

\rightline{\vbox{
                  \hbox{UTHEP-258}
                  \hbox{KUNS-1204}
                  \hbox{HE(TH)93/05}
		  \hbox{June 1993}
		  \hbox{\hep-th/9306067}
                  } }
\vskip 24pt
\centerline{
\fourteenrm
\baselineskip=20pt
\vbox{
\hbox{
Anomaly through Gauge Invariant Regularization }
\hbox{
with Infinite Number of Pauli-Villars Fields }
}}
\vskip 36pt
\centerline{
Sinya Aoki \footnote{*}{ saoki@ph.tsukuba.ac.jp } }
\vskip 12pt
\centerline{\it Institute of Physics, University of Tsukuba}
\centerline{\it Tsukuba, Ibaraki 305, Japan}
\vskip 16pt

\centerline{
Yoshio Kikukawa \footnote{**}{kikukawa@gauge.scphys.kyoto-u.ac.jp} }
\vskip 12pt
\centerline{\it Department of Physics, Kyoto University}
\centerline{\it Kyoto 606, Japan}

\vskip 32pt
\noindent
\centerline{\bf Abstract}
Abelian anomaly is examined by means of the recently proposed
gauge invariant regularization for SO(10) chiral gauge theory
and its generalization for a theory of arbitrary
gauge group with anomaly-free chiral fermion contents.
For both cases it is shown that the anomaly with correct normalization
can be obtained in a gauge invariant form without any counterterms.
\par
{\leftskip 0.4 in \rightskip 0.4 in
\par}

\vfill
\eject

\lref\tv{ G.~t'Hooft and M.Veltman, \NPB{44}{1972}{189}.}
\lref\adler{  S.~L.~Adler, \PR{177}{1969}{2426}. }
\lref\bell{  J.~S.~Bell and R.~Jackiw, \NC {60A}{1969}{47}. }
\lref\thooft{ G.~t'Hooft, \PRL{37}{1976}{110}; \PRD{14}{1976}{3432}.}
\lref\frolov{ S.~A.~Frolov and A.~A.~Slavnov, Saclay, SPhT/92-051. }
\lref\pauli{  W.~Pauli and F.~Villars, \RMP{21}{1949}{434}. }
\lref\baryon{ S.~Aoki, preprint UTHEP-236, Int. Journal of Modern Physics A
(in press).}
\lref\neuberger{ R.~Narayanan and H.~Neuberger, \PLB{303}{1993}{62}. }
\lref\slavnov{ L.~D.~Faddeev and A.~A.~Slavnov,
   {\it Gauge Fields. Introduction to Quantum Theory}, 2nd Edition,
    Benjamin, 1989.      }

\newsec{Introduction}
For chiral gauge theory any prescription of gauge invariant
regularization has not been known.
Dimensional regularization\tv,
although
it is a convenient and popular method for gauge theory,
does not preserve chiral invariance and thus gauge invariance
due to the nature of $\gamma_5$ in $4-2\epsilon$ dimensions.
This kind of explicit breaking of global chiral symmetry by regularizations
can not be avoided since its existence is necessary to reproduce
the abelian anomaly which has physical
meanings{\adler,\bell,\thooft}.
On the other hand, a possibility of a {\it chiral gauge invariant}
regularization has not been ruled out when the theory to be regularized is
{\it anomaly-free}.

Recently Frolov and Slavnov have proposed a method of gauge invariant
regularization for the fermion field in the irreducible spinor
(complex sixteen dimensional) representation of
SO(10) chiral gauge theory\frolov, which is automatically anomaly-free.
The method is based on the use of infinite number of
the Pauli-Villars fields\pauli\ and
it is shown that the chiral fermion determinant is made finite
in perturbation theory.
It also works for the anomaly-free set of fermions in the standard model
since these fermions combined with a right-handed neutrino
can be unified into the spinor representation of SO(10).

It is also noted that this regularization scheme has been considered
from more general point of view by Narayanan and Neuberger\neuberger.
They claimed that this method works for any anomaly free chiral gauge
theores.

Although these authors have shown that their methods
actually work for chiral fermion loops, there remains
several questions about it.
One of them is the question about abelian anomaly.
Is it possible to reproduce the abelian anomaly in
gauge invariant form by this method?
When we use the t'Hooft-Veltman definition
of $\gamma_5$ in the dimensional regularization, the chiral anomaly as well
as the U(1) Noether current is not gauge invariant\baryon.
We have to modify the current by adding the gauge non-invariant local
counterterm in order to obtain the gauge invariant form of the anomaly.
On the other hand, if we use the anti-commuting $\gamma_5$ to keep
gauge invariance, the chiral anomaly can not be obtained.
If the proposed method indeed regularize chiral fermions in gauge
invariant manner, we should obtain the gauge invariant form of the chiral
anomaly with correct normalization
(which is one-half of the anomaly in vector-like gauge theory),
without any gauge non-invariant counterterms for the current.

To clarify this point we will calculate explicitly the
abelian anomaly through the regularization procedure with infinite
number of the Pauli-Villars fields.
Before doing so, we will perform explicit calculations
of one loop diagrams and obtain the normalization of divergent
parts and finite contributions. We will also give a rule in order to
make the parity odd divergent parts vanish automatically (i.e.
by imposing the Bose symmetry), which appear in the generalization
for arbitrary gauge group.
This may be an important check to see whether the regularization is
gauge invariant.

\newsec{Regularized lagrangian of SO(10) chiral gauge theory}
Following Frolov and Slavnov, we consider a SO(10) gauge theory with
right-handed fermions which form an irreducible(16 dimensional) spinor
representation of SO(10). They are free from gauge anomaly.
For the regularization of fermion loops,
infinite number of the Pauli-Villars fields are introduced. They
consist of fermionic and bosonic right-handed spinors in
reducible(32 dimensional) spinor representations of SO(10).
As for the gauge fields, the higher covariant derivative
regularization\slavnov \ is adopted.

To consider the spinor representation of SO(10), let us introduce
thirty-two by thirty-two hermitian matrices
$\Gamma_i \; (i=1 \cdots 11)$ satisfying
$ \big\{ \Gamma_i, \Gamma_j \big\} = 2 \delta_{ij}$, and
$\big\{ \Gamma_i, \Gamma_{11} \big\} = 0$ ($i\not= 11$).
In terms of the $\Gamma_i$'s, SO(10) generators can be defined by
$ \sigma_{ij} \equiv {1\over2i} \Big[ \Gamma_i, \Gamma_j \Big],
(i,j=1 \cdots 10)$.
We use the notation $ T_a = \sigma_{ij}$ ( $ a= (i,j) $ ) for
simplicity and later generalization.
There exists a charge conjugation matrix $C$ such that
$C \Gamma_i^T C^{-1} = - \Gamma_i $ and $C T_a^T C^{-1} = - T_a $,
and therefore we have $\Tr ( T_a \{T_b, T_c\}) =0$.
This is why the spinor representation is anomaly-free.
It is also noted that $ C = C^{-1} = - \bar C = C^T $.
Let $ \psi_{+}$ denote the right-handed Weyl fermions
in the sixteen-dimensional irreducible spinor representation
of SO(10),
\eqn\spinor{ \psi_{+} = P_R \Gamma_{+} \psi_{+}, }
where
$ P_R = (1+\gamma_5)/2$ and $\Gamma_{\pm} = (1\pm\Gamma_{11})/2$.
Then the lagrangian of the theory is given by
\eqn\Lag{
{\cal L}_0 = \bar\psi_{+} i \fsm D \, \psi_{+}
-{1\over4}F_{\mu\nu}^{a}F^{\mu\nu a} ,
}
where $D_\mu = \partial_\mu -ig T_a A_\mu^a$.

For regularization we introduce
fermionic (anti-commuting)
and bosonic (commuting)
Pauli-Villars fields of right-handed chirality
in thirty-two-dimensional reducible spinor representation of SO(10)
such as
\eqn\PVf{ \psi_r = \psi_{r+} \oplus \psi_{r-}, \quad \psi_{r \pm} = P_R
\Gamma_\pm \psi_{r \pm}\quad (r=2,4, \dots,\infty), }
and
\eqn\PVb{ \phi_s = \phi_{s+} \oplus \phi_{s-}, \quad \phi_{s \pm } = P_R
\Gamma_\pm \phi_{s \pm}\quad (s=1,3, \dots, \infty). }
The mass terms for these regulator fields are given gauge-invariantly
as follows.
\eqn\LPV{ \eqalign{
{\cal L}_{PV}
&=\sum_r^\infty \bigl\{
 \bar\psi_{r} i \fsm D \, \psi_{r}
 -{M_r\over2}\Big( \bar\psi_{r} C_D \Gamma_{11} C\bar\psi_{r}^T
                 +\psi_{r}^T C_D C\Gamma_{11}\psi_{r} \Big)
\bigr\}
\cr
&+ \sum_s^\infty \bigl\{
\bar\phi_{s} i \fsm D \, \Gamma_{11} \phi_{s}
 -{M_s\over2}\Big( \bar\phi_{s} C_D C \bar\phi_{s}^T
                 +\phi_{s}^T C_D C \phi_{s} \Big)
\bigr\}
\cr
}
}
where $C_D$ is the usual charge conjugation matrix for Dirac fields, which
satisfies $C_D \gamma_\mu^T C_D^{-1}=-\gamma_\mu$ and
$C_D = \bar C_D = - C_D^T = - C_D^{-1}$.
Thus the regularized lagrangian of the theory consists of the sum
${\cal L}_0 + {\cal L}_{PV} $.
As can be seen later, the choice of $M_n$ such that $M_n = \Lambda n$ is
enought to regularize all fermion one-loop diagrams, where $\Lambda$
is a cut-off
in this regularization.
Weaker dependence of $M_n$ on $n$ such that $M_n=\Lambda \sqrt{n}$
is enough to regularize only the logarithmic divergences\neuberger.

To see how this method works as a gauge invariant regularization
we first consider fermion one loop contribution to two-point
function of gauge field in detail, which has been
discussed originally in ref.\frolov.
The contributions of $\psi_{+}$, $\psi_r$ and $\phi_s$
can be written as follows.
\eqn\twopointG{\eqalign{
\Gamma_{\mu\nu}^{ab}(p)
& \ = -g^2
  \int {d^4l \over (2\pi)^4}
{ \Tr( \gamma_\mu P_R T_a\Gamma_{+}
  (\fs l + \fs p) \gamma_\nu T_b \fs l )
 \over (l+p)^2 l^2}     \cr
&\quad \  - \sum_r g^2
  \int {d^4l \over (2\pi)^4}
{  \Tr( \gamma_\mu P_RT_a (\fs l + \fs p) \gamma_\nu T_b \fs l )
 \over [(l+p)^2-M_r^2][l^2-M_r^2]}     \cr
&\quad \  +\sum_s g^2
  \int {d^4l \over (2\pi)^4}
{\Tr( \gamma_\mu P_R T_a (\fs l + \fs p) \gamma_\nu T_b \fs l )
 \over [(l+p)^2-M_s^2][l^2-M_s^2]}     \cr
&\quad \  - \sum_r g^2
  \int {d^4l \over (2\pi)^4}
{ \Tr( \{\gamma_\mu P_RT_a\}^T M_r C C_D \Gamma_{11}
      \gamma_\nu T_b M_r C C_D \Gamma_{11}   )
 \over [(l+p)^2-M_r^2][l^2-M_r^2]}     \cr
&\quad  \ + \sum_s g^2
  \int {d^4l \over (2\pi)^4}
{ \Tr( \{\gamma_\mu P_R T_a\Gamma_{11} \}^T M_r C C_D
      \gamma_\nu T_b \Gamma_{11} M_r C C_D   )
 \over [(l+p)^2-M_s^2][l^2-M_s^2]}.     \cr
}}
To make the above expression finite,
we first sum the infinite number of contributions inside the momentum
integration.
It is noted that we have assigned the loop momenta
in the same way for all fields, $\psi_+$, $\psi_r$ and $\phi_s$.
This is a condition assumed implicitly to keep gauge invariance
in this regularization.
By this summation
we can define the two-point function with cutoff as
\eqn\Gtwodef{
\Gamma_{\mu\nu}^{ab}(p) {\buildrel {\rm def} \over = }
 - g^2  \Tr_+(T_a T_b )
\ \tilde \Gamma_{[R]\mu\nu}(p,0),
}
where $\Tr_+ (\cdot) = {1\over 2}\Tr(\cdot)$, and
for each matrix $A= P_R([R]),P_L ([L]),1([1])$ or $\gamma_5([5])$
we define
\eqn\GAdef{\eqalign{
\tilde \Gamma_{[A]\mu\nu}(p,0)
& \equiv
\int {d^4l \over (2\pi)^4}
\tr( \gamma_\mu A(\fs l + \fs p) \gamma_\nu \fs l )
\sum_{n=-\infty}^\infty {(-1)^n  \over [(l+p)^2-M_n^2][l^2-M_n^2]}     \cr
&
\hskip 1.2cm + \int {d^4l \over (2\pi)^4}
\tr( \gamma_\mu A \gamma_\nu )
\sum_{n=-\infty}^\infty {(-1)^n M_n^2  \over [(l+p)^2-M_n^2][l^2-M_n^2]}  \cr
}}
For $M_n=\Lambda n$
it is not hard to see the convergence of the momentum integration if
we refer to the following formul\ae\ and their asymptotic behaviors.
\eqn\Fdef{\eqalign{
F(\kappa^2)
&\equiv \sum_{n=-\infty}^\infty {\displaystyle (-1)^{n} \over
[\kappa^2 + n^2]}
={\pi \over \kappa}{1 \over \sinh(\pi\kappa)},   \cr
{\partial \over \partial \kappa^2} F(\kappa^2)
&= -\sum_{n=-\infty}^\infty {\displaystyle (-1)^n \over [\kappa^2 + n^2]^2}
=-{\pi \over 2\kappa^3}{1 \over \sinh(\pi\kappa)}
-{\pi^2 \over 2\kappa^2}{\cosh(\pi\kappa) \over \sinh^2(\pi\kappa)},  \cr
F_M(\kappa^2)
&\equiv \sum_{n=-\infty}^\infty {\displaystyle (-1)^n n^2\over
[\kappa^2 + n^2]^2}
=F(\kappa^2) + \kappa^2  F^\prime(\kappa^2)  \cr
&={\pi \over 2\kappa}{1 \over \sinh(\pi\kappa)}
-{\pi^2 \over 2}{\cosh(\pi\kappa) \over \sinh^2(\pi\kappa)},  \cr
}}
\eqn\Fasympt{
F(\kappa^2) , F^\prime(\kappa^2), F_M(\kappa^2)
\longrightarrow \kappa^{-n} \exp (-\pi\kappa)
\quad{\rm as  } \quad \kappa\rightarrow\infty \qquad (n=1,3 ,0).
}
Since the two-point function is finite at $\Lambda\not=\infty$,
any manipulations such as
the evaluation of the trace of gamma matrices, the introduction of
parameter integral and the change of momentum variables are
allowed. We can perform the momentum integral using Feynman
parameter integral and Wick rotation such as
$l^2 = - l_E^2$ and $\int d^4l = i\int d^4l_E$:
\eqn\Gtwocal{\eqalign{
&\tilde \Gamma_{[R]\mu\nu}(p,0) \cr
&=  4i (p^2\eta_{\mu\nu}-p_\mu p_\nu) \int {d^4l_E \over (2\pi)^4}
\int^1_0 dx x(1-x)
\sum_{n=-\infty}^\infty {(-1)^n \over [l_E^2+U+M_n^2]^2}
\cr
&\hskip 0.2cm +i \eta_{\mu\nu} \int {d^4l_E \over (2\pi)^4} \int^1_0 dx
\sum_{n=-\infty}^\infty  \Big(
{2 (-1)^n\over [l_E^2+U+M_n^2]}
-{l^2 (-1)^n\over [l_E^2+U+M_n^2]^2}
     \Big)
\cr
&= -{i\over 4\pi^2} (p^2\eta_{\mu\nu}-p_\mu p_\nu)
\int^1_0 dx x(1-x)
\int^\infty_0 d \kappa^2 \kappa^2
F^\prime \big(\kappa^2+u \big)
\cr
&\hskip 0.2cm +{i\over 16\pi^2} \eta_{\mu\nu} \Lambda^2 \int dx d\kappa^2
\Big[ (\kappa^2)^2 F^\prime\big(\kappa^2+u\big)
      +2 (\kappa^2) F\big(\kappa^2+u\big) \Big] \cr
&= -{i\over 4\pi^2} (p^2\eta_{\mu\nu}-p_\mu p_\nu)
\int^1_0 dx x(1-x)
2\log \Big[ \tanh \big({\pi \over 2\Lambda} \sqrt{-x(1-x) p^2} \,
                                      \big)  \Big]
\cr
&= -{i\over 4\pi^2} (p^2\eta_{\mu\nu}-p_\mu p_\nu)
\Big[\int^1_0 dx x(1-x) \log \big(-x(1-x) { p^2
\over \Lambda^2} \big)
+ {1\over3} \log(\pi/2) \Big]
\cr
}}
where $U= -x(1-x)p^2$ and $u=U/\Lambda^2$.
The gauge non-invariant piece proportional to $\eta_{\mu\nu}$,
which is quadratically divergent,
turns out to be zero after the integration by part in right-hand side of
the second equality and we finally obtain the gauge invariant result
\eqn\Gtworesult{ \eqalign{
\Gamma_{\mu\nu}^{ab}(p)
&=  g^2  \Tr_+(T_a T_b)
{i\over 24\pi^2} (p^2\eta_{\mu\nu}-p_\mu p_\nu)
\Big[ \log \big({ -p^2 \over \Lambda^2} \big) -{5\over 3}
+ 2 \log({\pi \over 2}) \Big]. \cr
}}
It is noted that the coefficient of the logarithmically divergent term is
equal to $\displaystyle {1\over 24\pi^2}$, which is one-half of the
Dirac fermion contribution in vectorial gauge theory.
Only one Weyl mode, not a Dirac mode ( = two Weyl modes),
contributes to the logarithmic
divergence in the vacuum polarization of gauge field and thus
to the beta function of gauge coupling constant. This shows that
the theory regularized by this method indeed describes a chiral gauge theory.
If $M_n =\Lambda \sqrt{n} $ is used instead of $M_n =\Lambda n $,
only the finite constant terms are modified such as
$ -{5\over 3}+2 \log({\pi\over 2})\rightarrow -{5\over 3}+\log({\pi\over 2})$.

Next we evaluate one loop contribution to the three-point function.
\eqn\threepointG{
\Gamma^{abc}_{\lambda\mu\nu}(p,k)
\equiv {\displaystyle g^3 \over 2}
\Big\{  \Tr_+(T_a T_bT_c)
       \tilde\Gamma_{[R]\lambda\mu\nu}(k+p,k,0)
       +\Tr_+(T_a T_cT_b)
       \tilde\Gamma_{[R]\lambda\nu\mu}(k+p,p,0) \Big\} ,
}
where for each matrix $A$ we define
\eqn\GthreeAdef{\eqalign{
&\tilde\Gamma_{[A]\lambda\mu\nu}(k+p,k,0) \cr
& \equiv \int {d^4l \over (2\pi)^4}
\tr \Big( \gamma_\lambda A (\fs l+\fs k+\fs p)
         \gamma_\mu (\fs l+\fs k)
         \gamma_\nu  \fs l        \Big)     \cr
& \hskip 20pt \times \sum_{n=-\infty}^\infty
{\displaystyle (-1)^n \over [(l+k+p)^2 -M_n^2][(l+k)^2-M_n^2][l^2-M_n^2]}
\cr
&+ \int {d^4l \over (2\pi)^4}
\tr \Big(
 \gamma_\lambda A (\fs l+\fs k+\fs p)\gamma_\mu\gamma_\nu
+ \gamma_\lambda A \gamma_\mu(\fs l+\fs k)\gamma_\nu
+ \gamma_\lambda A \gamma_\mu\gamma_\nu \fs l    \Big)    \cr
& \hskip 20pt
\times \sum_{n=-\infty}^\infty
{\displaystyle (-1)^2 M_n^2
\over [(l+k+p)^2 -M_n^2][(l+k)^2-M_n^2][l^2-M_n^2]} .  \cr
}
}
The momentum integrations in $\tilde\Gamma_{[A]\lambda\mu\nu}$
are also convergent for $M_n =\Lambda n$.
By the shift of integral variables and
the charge conjugation transformation, we can show an identity such
that
\eqn\RtoL{ \tilde\Gamma_{[R]\lambda\nu\mu}(k+p,p,0)
= -\tilde\Gamma_{[L]\lambda\mu\nu}(k+p,k,0)  }
and we have
\eqn\Gthree{\eqalign{
\Gamma^{abc}_{\lambda\mu\nu}(k,p)
& = {\displaystyle g^3 \over 2} \Tr_+(T_a [T_b, T_c])
\ {1\over2}\tilde\Gamma_{[1]\lambda\mu\nu}(k+p,k,0) \cr
& \hskip .5cm
+ {\displaystyle g^3 \over 4} \Tr_+(T_a \{T_b, T_c\})
\ {1\over2}\tilde\Gamma_{[5]\lambda\mu\nu}(k+p,k,0) . \cr
}}
Therefore the ``parity-odd'' term vanishes due to the anomaly-free
condition
\eqn\AnomalyC{\Tr_+(T_a \{T_b, T_c\})=0}
which is satisfied for the spinor
representation of SO(10).
Divergence of the ``parity-even'' part is easily shown to be
\eqn\Gthreecal{
{1\over 2}(k+p)^\lambda \tilde\Gamma_{[1]\lambda\mu\nu}(k+p,k,0)
= \tilde\Gamma_{[R]\mu\nu}(k,0) - \tilde\Gamma_{[R]\mu\nu}(p,0)
}
and we finally obtain the Ward-Takahashi identity
\eqn\Gthreeresult{
(k+p)^\lambda \Gamma^{abc}_{\lambda\mu\nu}(k,p)
 = {\displaystyle g^3 \over 2} \Tr_+ (T_a [T_b, T_c])
\ \Big( \tilde\Gamma_{[R]\mu\nu}(k,0) - \tilde\Pi_{[R]\mu\nu}(p,0) \Big).
}

Four point function is given by
\eqn\Gfourdef{\eqalign{
\Gamma^{abcd}_{\lambda\mu\nu\alpha}(p,k,q)
&= {\displaystyle g^4 \over 2}
\Big\{  \Tr_+(T_a T_bT_cT_d)
       \tilde\Gamma_{[R]\lambda\mu\nu\alpha}(q+k+p,q+k,q,0) \cr
    &   +\Tr_+(T_a T_dT_cT_b)
       \tilde\Gamma_{[R]\lambda\alpha\nu\mu}(q+k+p,p+k,p,0) \Big\} , \cr
}}
where $\tilde\Gamma_{[A]\lambda\mu\nu\alpha}$ is defined in an
analogous way
as before.
It is not difficult to see that the ``parity-odd'' term is proportional to
$\Tr_+( T_a T_b T_c T_d + T_d T_c T_b T_a)$.
It reduces to
$ i \Tr_+(f^{abe} T_e \{T_c,T_d\}+ f^{cde} T_e \{T_a,T_b\})$
where $f^{abc}$ is the structure constant defined by
$[T_a,T_b] =if^{abc} T_c$. Therefore the ``parity-odd' term again vanishes.

\newsec{Vector-like formulation and its extension to an arbitrary
gauge group}

The regularization in the previous section works for SO(10)
chiral gauge theory. In this section,
following Narayanan and Neuberger\neuberger,
we consider the generalization of the method for the case of
an arbitrary gauge group and find necessary conditions that the
regularization works.
We first rewrite
the Pauli-Villars Fields, $\psi_r$ and $\phi_s$,
in terms of fermionic and bosonic {\it Dirac} fields.
Introducing formal notations $\Psi_R$ and
$\Psi_L$ as right- and left- handed fermionic fields
with infinite number of components,
\eqn\DiracPVf{
\Psi_{R}
\equiv \pmatrix{ \psi_{+}  \cr
                 \psi_{2+} \cr
                 \psi_{4+} \cr
                 \vdots    }, \qquad
\Psi_{L}
\equiv \pmatrix{  CC_D\bar\psi_{2-}^T \cr
                  CC_D\bar\psi_{4-}^T \cr
                   \vdots
        } ,
}
and also $\Phi_R$ and $\Phi_L$ for bosonic fields,
\eqn\DiracPVb{
\Phi_{R}
\equiv \pmatrix{ \phi_{1+}  \cr
                 \phi_{3+} \cr
                 \phi_{5+} \cr
                 \vdots    }, \qquad
\Phi_{L}
\equiv \pmatrix{  CC_D\bar\phi_{1-}^T \cr
                  CC_D\bar\phi_{3-}^T \cr
                  CC_D\bar\phi_{5-}^T \cr
                   \vdots
        } ,
}
the lagrangian can be rewritten as follows.
\eqn\Lvec{\eqalign{
{\cal L}_{VEC}
& = \bar\Psi i \fsm D \Psi
       - \left( \bar\Psi_L M_\infty \Psi_R
               + \bar\Psi_R M_\infty^\dagger \Psi_L \right) \cr
& + \bar\Phi i \fsm D \Phi
       - \left( \bar\Phi_L M_\infty^\prime \Phi_R
               + \bar\Phi_R M_\infty^{\prime\dagger} \Phi_L \right), \cr
}
}
where the mass matrices are defined as
\eqn\Massmatrix{
M_\infty  = \pmatrix{
 0      & 2M     & 0      & 0      & \ldots  \cr
 0      & 0      & 4M     & 0      & \ldots   \cr
 0      & 0      & 0      & 6M     & \ldots  \cr
 \vdots & \vdots & \vdots & \vdots & \ddots
}, \quad
M_\infty^\prime  = \pmatrix{
 M      & 0      & 0      & 0      & \ldots  \cr
 0      & 3M     & 0      & 0      & \ldots   \cr
 0      & 0      & 5M     & 0      & \ldots  \cr
 \vdots & \vdots & \vdots & \vdots & \ddots
}.
}
It is noted that all fields in $\Psi$ and $\Phi$ are now in the
{same} sixteen-dimensional irreducible spinor representation of
SO(10).
The chiral nature of the theory is ensured by the condition
\eqn\Chiralcondition{
 {\rm dim } \big[ {\rm Ker}( M_\infty^\dagger M_\infty) \big]
- {\rm dim } \big[ {\rm Ker} ( M_\infty M_\infty^\dagger )\big] = 1.
}

In this formulation it is obvious that
the ``parity-odd'' contributions of the Pauli-Villars Fields are absent.
Therefore the ``parity-odd'' part of the original fermion is not
regularized\neuberger. 
As we have shown explicitly in the previous section, however, such parity-odd
parts in loop integrals does vanish because of the bose symmetry
and the anomaly cancellation condition because
they are regularized finitely.
Actually we can argue even in the vector-like formulation
that these parity-odd contributions are to vanish.
To do this we adopt the following rule.
Consider n-point function of external
gauge field in the fixed order such that
$A_{\mu_1}^{a_1}$, $A_{\mu_2}^{a_2}$, $\cdots$, $A_{\mu_n}^{a_n}$.
Then there always exist two types of diagrams. In one of them
{\it n\/} fermion lines form a loop in one direction such
that $1\rightarrow 2\rightarrow\cdots \rightarrow n
\rightarrow 1$. In the other
they form a loop in the opposite direction such that
$1\rightarrow n\rightarrow n-1\rightarrow\cdots \rightarrow 2\rightarrow 1$.
The rule we adopt here is that
we assign the  momentum of fermion line between
$A_{\mu_1}^{a_1}$ and $A_{\mu_n}^{a_n}$  {\it same} in  both diagrams.
By this rule we obtain the following expressions for
two-, three- and four-point functions:
\eqn\Gvec{\eqalign{
\Gamma_{\mu\nu}^{ab}(p) & = -{g^2\over 4}
[ \Tr_+ (T_a T_b) \left\{\tilde\Gamma_{[1]\mu\nu}(p,0)
+\Gamma^{0}_{[5]\mu\nu}(p,0)\right\} \cr
 & + \Tr_+ (T_a T_b)
\left\{\tilde\Gamma_{[1]\mu\nu}(0,-p)+\Gamma^{0}_{[5]\mu\nu}(0,-p)\right\}
] \cr
\Gamma_{\lambda\mu\nu}^{abc}(p,k) & = {g^3\over 4}
[ \Tr_+ (T_a T_b T_c) \left\{\tilde\Gamma_{[1]\lambda\mu\nu}(k+p,k,0)
+\Gamma^0_{[5]\lambda\mu\nu}(k+p,k,0) \right\} \cr
& + \Tr_+ (T_a T_b T_c) \left\{\tilde\Gamma_{[1]\lambda\nu\mu}(0,-k,-k-p)
+\Gamma^0_{[5]\lambda\nu\mu}(0,-k,-k-p)\right\}] \cr
\Gamma_{\lambda\mu\nu\alpha}^{abcd}(p,k,q) & = -{g^4\over 4}
[ \Tr_+ (T_a T_b T_c T_d)
\left\{\tilde\Gamma_{[1]\lambda\mu\nu\alpha} (q+k+p,q+k,q,0)\right. \cr
&
\qquad +\left. \Gamma^0_{[5]\lambda\mu\nu\alpha} (q+k+p,q+k,q,0)\right\} \cr
& + \Tr_+ (T_a T_d T_c T_b)\left\{
\tilde\Gamma_{[1]\lambda\alpha\nu\mu}(0,-q,-q-k,-q-k-p)\right. \cr
& \qquad
+\left. \Gamma^0_{[5]\lambda\alpha\nu\mu}(0,-q,-q-k,-q-k-p) \right\}] \cr
}
}
where $\tilde \Gamma_{[1]\mu\nu}$, $\tilde \Gamma_{[1]\lambda\mu\nu}$
and $\tilde \Gamma_{[1]\lambda\mu\nu\alpha}$ are
parity-even parts of two-, three- and four-point functions
calculated in the Weyl notation of the previous section.
The unregularized parity-odd terms are given by
\eqn\Parityodd{\eqalign{
\Gamma^0_{[5]\mu\nu}(p,0) &=  \int {d^4l \over (2\pi)^4}
Tr( \gamma_\mu \gamma_5 (\fs l + \fs p) \gamma_\nu \fs l )
{1 \over (l+p)^2 l^2}     \cr
\Gamma^0_{[5]\lambda\mu\nu}(p_1,p_2,0) & =  \int {d^4l \over (2\pi)^4}
{ \Tr \Big( \gamma_\lambda \gamma_5 (\fs l+\fs p_1)
         \gamma_\mu (\fs l+\fs p_2)
         \gamma_\nu  \fs l        \Big)
 \over (l+p_1)^2 (l+p_2)^2 l^2 }
\cr }
}
\eqn\GfourParityodd{
\Gamma^0_{[5]\lambda\mu\nu\alpha}(p_1,p_2,p_3,0)  =
\int {d^4l \over (2\pi)^4}
{ \Tr \Big( \gamma_\lambda \gamma_5 (\fs l+\fs p_1)
         \gamma_\mu (\fs l+\fs p_2)
         \gamma_\nu (\fs l+\fs p_3)
         \gamma_\alpha  \fs l        \Big)
 \over (l+p_1)^2 (l+p_2)^2 (l+p_3)^2 l^2 } .
}
Taking transpose of the Dirac matrices in the traces and changing
the integration variables $l$ to $-l$, we find that
\eqn\GvecCal{\eqalign{
\Gamma_{\mu\nu}^{ab}(p) & = -{g^2\over 4}
 \Tr_+ (T_a T_b) \tilde\Gamma_{[1]\mu\nu}(p,0) \cr
\Gamma_{\lambda\mu\nu}^{abc}(p,k) & = {g^3\over 4}
\left[ \Tr_+ (T_a [T_b, T_c]) \tilde\Gamma_{[1]\lambda\mu\nu}(k+p,k,0)
\right.\cr
& \qquad + \left.
\Tr_+(T_a \{T_b,T_c\})\Gamma^0_{[5]\lambda\mu\nu}(k+p,k,0) \right] \cr
\Gamma_{\lambda\mu\nu\alpha}^{abcd}(p,k,q) & = -{g^4\over 4}
\left[ Tr_+ (T_a T_b T_c T_d+T_aT_dT_cT_b)\tilde\Gamma_{[1]\lambda\mu\nu\alpha}
(q+k+p,q+k,q,0)\right. \cr
& + \left. \Tr_+(T_aT_bT_cT_d-T_aT_dT_cT_b)\Gamma^0_{[5]\lambda\mu\nu\alpha}
(q+k+p,q+k,q,0) \right] .
}}
Now it is easy to see that the unregularized parity-odd contributions vanish
due to the group structure of SO(10) as mentioned before.
The remaining parity-even contributions agree with those
calculated in the previous section.

Now we discuss the extension of this regularization method in the
{\it vector-like} formulation to an arbitrary gauge group.
With the help of same rule as above,
we can show that unregularized parity-odd terms are
proportional to $\Tr_f ( T_a \{T_b,T_c\})$ or
$\Tr_f (T_aT_bT_cT_d-T_aT_dT_cT_b)$,
where $\Tr_f$ means trace over a given representation $f$
of the gauge group.  Since $\Tr_f (T_a \{T_b,T_c\}) \not= 0$
for the complex representation $f$,
this regularization gives neither finite nor gauge invariant results
at finite cut-off.
Only when we introduce several fermion {\it flavors}
such that
$ \sum_f \Tr_f (T_a \{T_b,T_c\}) = 0$ which also implies
$ \sum_f \Tr_f (T_aT_bT_cT_d-T_aT_dT_cT_b)= 0$,
this regularization gives finite and gauge-invariant results.
Therefore this regularization in the vector-like formulation
works only for {\it anomaly-free} fermion contents but of general
gauge group, 
as claimed in ref.\neuberger.

\newsec{Chiral Anomaly}

The lagrangian ${\cal L}_0$ in consideration has a U(1) symmetry under the
transformation:
\eqn\ChiralTrans{
\psi_{+} \longrightarrow \exp\{i\alpha\} \, \psi_{+} .
}
The associated current is given by
\eqn\NetherCurrent{ \bar\psi_{+} \gamma_\mu \psi_{+}.
}
It is well known that this current suffers from the anomaly,
which would have physical meanings.

To evaluate the anomaly we must regularize the above current.
We find that the following definition of current is regularized and
gauge invariant:
\eqn\Current{\eqalign{
J^5_\mu
& = \bar\psi_{+} \gamma_\mu \psi_{+}
+\sum_r \bar\psi_{r} \gamma_\mu \psi_{r}
+\sum_s \bar\phi_{s} \gamma_\mu \Gamma_{11} \phi_{s}  \cr
}
}
in the Weyl (chiral) notation, or
\eqn\CurrentVect{\eqalign{
J^5_\mu
& = \bar\Psi \gamma_\mu\gamma_5 \Psi
+ \bar\Phi \gamma_\mu \gamma_5 \Phi  \cr
}
}
in the Dirac(Vector-like) notation.
This current is the Noether current associated with the transformation
\eqn\WeylChiralTrans{\left\{
\eqalign{
\psi_{+} & \longrightarrow \exp\{i\alpha\} \, \psi_{+} \cr
\psi_{r} & \longrightarrow \exp\{i\alpha\} \, \psi_{r} \cr
\phi_{s} & \longrightarrow \exp\{i\alpha\} \, \phi_{s} \cr
} \right.}
in the Weyl notation, or
\eqn\DiracChiralTrans{\left\{
\eqalign{
\Psi & \longrightarrow \exp\{i\alpha\gamma_5\} \, \Psi \cr
\Phi & \longrightarrow \exp\{i\alpha\gamma_5\} \, \Phi \cr
} \right.}
in the Dirac notation.
We calculate two-point function of gauge field with the current
$J^5_\mu $ inserted, $\Gamma^{5 ab}_{\lambda\mu\nu}(k,p)$,
according to the rules in the previous two sections.
At one-loop we find
\eqn\AnomalyVertexDef{
\Gamma^{5 ab}_{\lambda\mu\nu}(p,k)
\equiv -i{\displaystyle g^2 \over 2} \Tr_+( T_aT_b)
\ \Big\{ \tilde\Gamma_{[R]\lambda\mu\nu}(k+p,k,0)
        +\tilde\Gamma_{[R]\lambda\nu\mu}(k+p,p,0) \Big\}
}
in the Weyl notation, or
\eqn\AnomalyVertexDefDirac{\eqalign{
\Gamma^{5 ab}_{\lambda\mu\nu}(p,k)
& \equiv -i{\displaystyle g^2 \over 4} \Tr_+( T_aT_b)
\ \Big\{ \tilde\Gamma_{[5]\lambda\mu\nu}(k+p,k,0)+
        \tilde\Gamma^0_{[1]\lambda\mu\nu}(k+p,k,0)   \cr
&        +\tilde\Gamma_{[5]\lambda\nu\mu}(0,-k, -k-p)
        +\tilde\Gamma^0_{[1]\lambda\nu\mu}(0,-k, -k-p) \Big\}
}}
in the Dirac notation, where $\Gamma_{[A]\lambda\mu\nu}$ and
$\tilde\Gamma^0_{[A]\lambda\mu\nu}$ are given in the previous sections.
Since
\eqn\Grelation{ \eqalign{
\tilde\Gamma_{[R]\lambda\nu\mu}(k+p,k,0)
& =  -\tilde\Gamma_{[L]\lambda\mu\nu}(k+p,p,0) \cr
\tilde\Gamma_{[5]\lambda\nu\mu}(k+p,k,0)
& =  \tilde\Gamma_{[5]\lambda\mu\nu}(0,-k,-k-p) \cr
\tilde\Gamma^0_{[1]\lambda\nu\mu}(k+p,k,0)
& = - \tilde\Gamma^0_{[1]\lambda\mu\nu}(0,-k,-k-p) \cr
}  }
we obtain
\eqn\AnomalyVertexCal{
\Gamma^{5 ab}_{\lambda\mu\nu}(p,q)
\equiv -i{\displaystyle g^2 \over 2} \Tr_+( T_aT_b)
 \tilde\Gamma_{[5]\lambda\mu\nu}(k+p,k,0)
}
in both Weyl and Dirac notations.

Divergence of this parity-odd term is evaluated as follows.
$$\eqalign{
&(k+p)^\lambda \tilde\Gamma_{[5]\lambda\mu\nu}(k+p,k,0)  \cr
&=- 8i \epsilon_{\rho\sigma\mu\nu} k^\rho p^\sigma
   \int {d^4l \over (2\pi)^4}
    \sum_{n=-\infty}^\infty
{\displaystyle (-1)^n M_n^2
\over [(l+k+p)^2 -M_n^2][(l+k)^2-M_n^2][l^2-M_n^2]}   \cr
& = - 8i \epsilon_{\rho\sigma\mu\nu} k^\rho p^\sigma
2 \int_0^1 dx \int_0^{1-x} dy
   \int {d^4l \over (2\pi)^4} \sum_{n=-\infty}^\infty
{\displaystyle (-1)^n M_n^2 \over [l^2 -V-M_n^2]^3}       \cr
}$$
where $V=-x(k+p)^2-yk^2+(x(k+p)+yk)^2$ and $v=V/\Lambda^2 $
\eqn\DivParityodd{
\eqalign{
 & = -{1\over 2\pi^2} \epsilon_{\rho\sigma\mu\nu} k^\rho p^\sigma
2 \int_0^1 dx \int_0^{1-x} dy
   \int {d\kappa^2 \; \kappa^2} \sum_{n=-\infty}^\infty
{\displaystyle (-1)^n n^2 \over [\kappa^2 +f+n^2]^3}
\cr
 & = {1\over 2\pi^2} \epsilon_{\rho\sigma\mu\nu} k^\rho p^\sigma
\times \cr
& \hskip 10pt \times 2 \int_0^1 dx \int_0^{1-x} dy
   \int {d\kappa^2 \; \kappa^2}
\Big( {\partial \over \partial \kappa^2}F(\kappa^2+f)
 +{\kappa^2 +f\over 2}
\left( {\partial \over \partial \kappa^2}\right)^2
F(\kappa^2+f) \Big)                                      \cr
 & = {1\over 2\pi^2} \epsilon_{\rho\sigma\mu\nu} k^\rho p^\sigma
2 \int_0^1 dx \int_0^{1-x} dy
{f\over 2} \int {d\kappa^2 {\partial \over \partial \kappa^2}}F(\kappa^2+f) \cr
 & = {1\over 4\pi^2} \epsilon_{\rho\sigma\mu\nu} k^\rho p^\sigma. \cr
}}

By similar calculation we obtain three-point function of gauge field
with the current $J_\mu^5$ inserted as
\eqn\AnomalyThreepoint{
\Gamma^{5 abc}_{\lambda\mu\nu\alpha}(p,q,k)
\equiv i{\displaystyle g^3 \over 2} \Tr_+( T_aT_bT_c)
 \tilde\Gamma_{[5]\lambda\mu\nu\alpha}(q+k+p,q+k,q,0)
}
in both Weyl and Dirac notations. Its divergence becomes
\eqn\AnomalyThreepointResult{
(q+k+p)^\lambda \tilde\Gamma_{[5]\lambda\mu\nu\alpha}(q+k+p,q+k,q,0)
= -{1\over 4\pi^2}\epsilon_{\rho\mu\nu\alpha}(q+p)^\rho
}

Combining the above two results, we obtain the vacuum expectation value
of $\partial^\mu J_\mu^5$ in the presence of back-ground gauge field
$A_\mu$ as follows.
\eqn\AnomalyResult{
< \partial^\mu J_\mu^5 >_{A_\mu}
= -{1\over 2}\times\left[{i g^2\over 16\pi^2} \epsilon_{\mu\nu\alpha\beta}
\Tr_{+} (F^{\mu\nu} F^{\alpha\beta}) \right] .
}
Thus we have obtained the gauge invariant form of
abselian anomaly with the correct normalization for
chiral fermion $\psi_{+}$,
without any gauge non-invariant local counterterms.

For an anomaly-free set of complex representations
of a general gauge group,
the calculation in the Dirac notation gives
the same result as eq.\AnomalyResult\
with the trace $\Tr_{+}$ replaced by the traces $\Tr_f$ summed
over flavors.

We can understand the above result by the following naive argument.
Since all fileds except one massless mode are massive Dirac fields,
each of them should produce a contribution to the divergence of
the current $J_\mu^5$. They are evaluated gauge invariantly like
in QCD. The normalization of the chiral anomaly amounts to the
infinite sum of fermionic (+1) and bosonic (-1) contributions.
With a suitable regularization, this can be equal to one half,
\eqn\inftysum{
-1 + 1 - 1 + 1 -1 + \cdots = - {1\over 2},
}
which is just the normalization for a Weyl (chiral) fermion.
The calculation in this section shows that this naive expectation is indeed
correct.

\newsec{Conclusion and discussion}

We have
investigated newly proposed gauge invariant regularization for
the SO(10) chiral gauge theory and its generalization to other chiral
gauge theories.

In the Weyl notation this method of regularization works when the
internal momenta are assigned in the same way for all
diagrams generated by different fields (the physical fermion and regulators).
In the Dirac notation we need more careful assignment of loop momenta.
Otherwise the finiteness breaks down and the gauge non-invariant
contribution may arise.
But once we adopt such rules for the assignment of momenta, the
condition of anomaly cancellation is enough for chiral gauge
invariance of regularized green functions.
Thus it can be shown
that this regularization
also works as gauge invariant regularization for
chiral gauge theory of an arbitrary gauge group with an anomaly-free
set of fermions\neuberger.

The main result of this paper is that
using this regularization
we have obtained
the gauge invariant form of chiral anomaly with the
normalization for a Weyl fermion. So far
this regularization is the first one which gives the
chiral anomaly gauge-invariantly in {\it chiral } gauge
theory\baryon.

As far as perturbation theory is concerned,
we now have at least one gauge invariant regularization scheme for
an arbitrary chiral gauge theory with anomaly-free fermion contents.
There may be a lot of applications of this regularization in future.

\vskip 30pt
\leftline{\bf Acknowledgement}\hfill\break
We would like to express our sincere thanks to T.~Kugo for
informing us of the work of S.~A.~Frolov and A.~A.~Slavnov.
We also appreciate discussions with T.~Yanagida and H.~Murayama.

\listrefs
\bye
\end